\documentstyle[prl,aps,twocolumn,epsf]{revtex}
\begin{document} 
\draft
\narrowtext 

\title{Stable characteristic evolution of generic 3-dimensional single-black-hole spacetimes}

\author{R.~G\'omez$^{\rm a}$, L.~Lehner$^{\rm a}$, R.~L.~Marsa$^{\rm a,b}$, J.~Winicour$^{\rm a}$ \\
        A.~M.~Abrahams$^{\rm c}$,
        A.~Anderson$^{\rm d}$, P.~Anninos$^{\rm c}$,
        T.~W.~Baumgarte$^{\rm c}$, N.~T.~Bishop$^{\rm e}$,
        S.~R.~Brandt$^{\rm c}$, J.~C.~Browne$^{\rm b}$,
        K.~Camarda$^{\rm f}$, M.~W.~Choptuik$^{\rm b}$, R.~R.~Correl$^{\rm b}$,
        G.~B.~Cook$^{\rm g}$, C.~R.~Evans$^{\rm d}$,
        L.~S.~Finn$^{\rm h}$, G.~C.~Fox$^{\rm i}$,
        T.~Haupt$^{\rm i}$, M.~F.~Huq$^{\rm b}$,
        L.~E.~Kidder$^{\rm h}$, S.~A.~Klasky$^{\rm i}$,
        P.~Laguna$^{\rm f}$,
        W.~Landry$^{\rm g}$, 
        J.~Lenaghan$^{\rm d}$, 
        J.~Masso$^{\rm c}$, R.~A.~Matzner$^{\rm b}$,
        S.~Mitra$^{\rm b}$, P.~Papadopoulos$^{\rm f}$,
        M.~Parashar$^{\rm b}$, L.~Rezzolla$^{\rm c}$,
        M.~E.~Rupright$^{\rm d}$, F.~Saied$^{\rm c}$,
        P.~E.~Saylor$^{\rm c}$, M.~A.~Scheel$^{\rm g}$,
        E.~Seidel$^{\rm c}$, S.~L.~Shapiro$^{\rm c}$,
        D.~Shoemaker$^{\rm b}$, L.~Smarr$^{\rm c}$,
        B.~Szil\'agyi$^{\rm a}$, S.~A.~Teukolsky$^{\rm g}$,
        M.~H.~P.~M.~van Putten$^{\rm g}$, P.~Walker$^{\rm c}$,
        J.~W.~York Jr$^{\rm d}$.}
\address{$^{\rm a}$University of Pittsburgh, Pittsburgh, Pennsylvania 15260}
\address{$^{\rm b}$The University of Texas at Austin,
                   Austin, Texas 78712}
\address{$^{\rm c}$University of Illinois at
                   Urbana-Champaign, Urbana, Illinois 61801}
\address{$^{\rm d}$University of North Carolina, Chapel Hill,
North Carolina 27599}
\address{$^{\rm e}$University of South Africa, P.O. Box 392,
                   Pretoria 0001, South Africa}
\address{$^{\rm f}$Penn State University, University Park, Pennsylvania 16802}
\address{$^{\rm g}$Cornell University, Ithaca, New York 14853}
\address{$^{\rm h}$Northwestern University, Evanston, Illinois 60208}
\address{$^{\rm i}$Syracuse University, Syracuse, New York 13244-4100}

\maketitle

\begin{abstract}
We report new results which establish that the accurate 3-dimensional
numerical simulation of generic single-black-hole spacetimes has
been achieved by characteristic evolution with unlimited long term
stability. Our results cover a selection of distorted, moving and
spinning single black holes, with evolution times up to $60,000M$.
\end{abstract}


Accurate numerical simulation of black holes is necessary to
calculate gravitational waveforms in the nonlinear regime that cannot
be approximated by perturbation theory. The importance of such
waveforms to the success of the LIGO gravity wave detector was a prime
factor in organizing the Binary Black Hole Grand Challenge Alliance,
whose goal is to provide the capability of obtaining waveforms from the
inspiral and merger of binary black holes~\cite{all}. The
Alliance is developing a code consisting of a Cauchy module matched to an outer boundary module using
either a characteristic or perturbative method.
For reports on the Cauchy and perturbative modules see~\cite{all1} and~\cite{all2}, respectively.
Here we report two new tests of a 3-dimensional characteristic
evolution module~\cite{high,wobb} that establish its unlimited
capability to accurately simulate a generic single black hole spacetime
and that establish a calibrated tool to attack the binary problem:
(i) We have evolved a black hole of mass M moving
with periodic time dependence induced by a coordinate wobble for a time
of $60,000M$; and (ii) we
have evolved an initially distorted, spinning black hole up to the
final equilibrium state, which remains stationary to within machine
round-off error and is a discretized version of the Kerr black hole 
spacetime.
(We use units with Newton's constant $G=1$ and the speed of light $c=1$.  
Thus $M \equiv GM/c^3$ is a time. Also, $M \equiv GM/c^2$ is a length.)  

In the 1970s and 1980s, the difficulty of stably simulating even a
strictly spherical (one spatial dimension) single black hole led to the
formulation of ``the Holy Grail of numerical relativity'', a list of
requirements for `` a code that simultaneously

	$\bullet$ Avoids singularities

	$\bullet$ Handles black holes

	$\bullet$ Maintains high accuracy

	$\bullet$ Runs forever.''~\cite{grail}

The results reported here definitely achieve this goal in the
3-dimensional, single black hole case. The challenge for the 1990s and
beyond is the Binary Black Hole problem. The results here may become
directly applicable to that {\it multiple} black hole stage.

The characteristic algorithm is a new computational treatment of
hyperbolic systems. The theoretical framework is the characteristic
initial value problem, pioneered by Bondi~\cite{bondi} and
Penrose~\cite{penrose} in the 1960s. Almost all numerical modeling of
hyperbolic systems has been based upon the Cauchy initial value
problem, which evolves fields on spacelike hypersurfaces along a
discrete sequence of time steps. The major new idea in the
characteristic approach is to evolve fields on outgoing (or ingoing)
light cones along a sequence of retarded (or advanced) time steps.
Figure~1 shows the schematic setup for the outgoing case. A world tube
$\Gamma$ has been placed as an inner boundary on the light cones
(characteristics) to excise caustics from the evolution domain.
Boundary data on $\Gamma$ and data on the initial light cone ${\cal
N}_0$ determine a unique exterior evolution.

Characteristic evolution has several advantageous
features~\cite{gwsouth}: The evolution variables reduce to one
complex function related to the two gravitational
polarization modes; the Einstein equations reduce to propagation
equations along the light rays; and there are no constraints on the
initial data. Furthermore, because the light cones are the spacetime
hypersurfaces along which waves propagate, such propagating
disturbances appear fairly smooth along them. This feature allows
implementation of Penrose's spacetime compactification~\cite{penrose}
to include points at future lightlike infinity (in the case based on
outgoing light cones), where the waveform is calculated in the
numerical grid. The major disadvantage is the difficulty in treating
caustics. One early strategy for a characteristic algorithm proposed
tackling the caustics head-on as part of the evolution~\cite{friedst2}.
But to date this has only been accomplished for point caustics in
axisymmetric spacetimes~\cite{papa}; and the extension to 3D would be
prohibitive on present-day machines.  Cauchy-characteristic matching is
a strategy for combining the complementary strengths of Cauchy and
characteristic evolution.

The implementation and calibration of the 3D characteristic module has
been described elsewhere~\cite{high,cce}. For a grid of discretization
size $\Delta$, the numerical solutions converge in the continuum limit
to exact analytic values in a wide variety of test beds, with
$O(\Delta^2)$ error. The long term stability of the {\it outgoing}
problem has also been established~\cite{high}. In these studies, the
inner world tube $\Gamma$ was chosen to be the ingoing branch of the
$r=2M$ horizon in a Schwarzschild spacetime. The initial data consisted
of a pulse of ingoing radiation on ${\cal N}_0$. This set the data for
the scattering of a pulse of radiation by a Schwarzschild black hole,
the classic problem first studied perturbatively by Price~\cite{price}.
The angular momentum of the ingoing pulse leads to a final black hole
with spin. The pulse is partially transmitted into the black hole and
partially scattered to (compactified) infinity along outgoing light
cones, where its waveform is obtained. The evolution handles highly
distorted black holes with backscattered radiation a thousand times
more massive than the initial black hole and with a peak power $\approx
10^5$ in dimensionless units (equivalent to conversion of our galaxy's
mass into gravitational waves in 1 second).

The calculation of the waveform at infinity
for the binary problem can be posed in a similar way by
matching a Cauchy interior module matched at a worldtube $\Gamma$
to a characteristic outer module. (See~\cite{all2} for
an alternative perturbative matching scheme). In model
3D nonlinear problems, Cauchy-characteristic matching dramatically
outperforms other existing outer boundary conditions for Cauchy
evolution~\cite{jcp97}. It has been successful in 1D general
relativity~\cite{dinv,excise} but its efficacy in 3D general
relativity is yet to be determined, because a stable matching 
scheme has not yet been found.

A simple transformation switches an {\it outgoing} characteristic
evolution module into an {\it ingoing} module~\cite{wobb,excise}. In
this case, to uniquely define a
black hole spacetime, boundary data is prescribed on an outer worldtube
and on an incoming light cone ($\Gamma$ and ${\cal N}_0$ in Fig.~2); and
in order to excise the singular region interior to the black hole, an
inner boundary is constructed at a world tube traced out by a {\it
marginally trapped surface} (${\cal T}$ in Fig.~2). This
extends to characteristic evolution the strategy initially
proposed by Unruh (see~\cite{thornburg1987}) for Cauchy evolution
of black holes.

This strategy is based upon the properties of {\it trapped
surfaces}~\cite{pentrap}. Normally, the light rays emitted in the
outward normal direction to a (topologically) spherical surface form an
expanding beam. But strong gravitational lensing can make such an
outgoing spherical beam everywhere convergent. Such a surface whose
outgoing and ingoing rays all converge is called trapped.  A {\it
marginally} trapped surface (MTS) is the borderline case in which the
outward light cone neither expands nor converges. Under reasonable
assumptions, a MTS cannot lie outside a black hole (see~\cite{wald}).
Consequently, if the worldtube $\Gamma$ in Fig.~2 is outside the black
hole then the ingoing light cone ${\cal N}_0$ must extend some finite
distance inward from $\Gamma$ before reaching a MTS (${\cal S}$ in
Fig.~2). In all known examples of black holes the singularities are
located inside a MTS. Excision of the interior of the MTS thus protects
the evolution from encountering a singularity.

In order to implement this strategy (i) the evolution module must be
equipped with an MTS finder and (ii) the singular region inside the MTS
must be excised from the computational grid without influencing the
exterior evolution. In a characteristic evolution, item (i) is
facilitated by locating the MTS in a natural way by deforming an
initial guess along the ingoing light rays~\cite{wobb}.  Similarly,
item (ii) is facilitated because the excision of the interior of the
MTS reduces to a 1-dimensional problem with respect to a radial grid
variable. There is no need for any further boundary condition on the
MTS: by construction, waves emitted from its surface cannot expand into
the exterior region. This theoretical property is built into the
characteristic algorithm.

Details and calibration of the {\it ingoing} module are given
in~\cite{wobb}. In initial simulations of a non-spinning black hole,
data on $\Gamma$ was induced from the exterior geometry of a
Schwarzschild spacetime and initial data on ${\cal N}_0$ consisted of a
Schwarzschild black hole of mass $M$ distorted by a pulse of radiation.
The worldtube $\Gamma$ was also placed in motion relative to the static
symmetry of the exterior Schwarzschild spacetime to produce a time
dependent location of the black hole in the numerical grid. The
dynamics was monitored by tracking the surface area ${\cal A}$ of the
MTS. For a non-spinning black hole in equilibrium, this surface area
equals its Schwarzschild value ${\cal A_S}= 16\pi M^2$. Calculation of
${\cal A}$ is an especially demanding test when the  world tube is
offset from the spherical symmetry of the Schwarzschild exterior and
then placed in a periodic circular orbit.  This periodic wobble of the
coordinates leads to a periodic time dependence of both the metric and
the location of the MTS, even in the final state of intrinsically
static equilibrium. The area of a MTS determines its Hawking
mass~\cite{hawkm} and gives a useful measure of the energy inside it.
Initially, ${\cal A}<{\cal A_S}$ due to the energy content of the
initial pulse on ${\cal N}_0$.  The MTS grows as this energy falls into
it. For a non-spinning black hole, ${\cal A}\rightarrow {\cal A_S}$  as
the MTS settles into equilibrium, even though the metric and the
location of the MTS vary periodically (see Fig.~3).

It is important to establish that the ingoing characteristic module has
{\it no long term instabilities} and that it can handle {\it spinning
black holes}.  We now present two new tests which demonstrate that it
can essentially evolve a generic black hole ``forever''.

Since ``forever'' cannot be rigorously attained in any finite
simulation, we appeal to a characteristic time necessary to obtain
accurate waveforms for the inspiral and merger of two black holes. If
one of the holes is small then the test particle approximation can be
used. Consider a test particle in a quasi-circular orbit about a
Schwarzschild black hole, where the final stable orbit is at $r=6M$.
From the quadrupole approximation (see~\cite{wald}), the radiation rate
per orbit is $\approx 10^{-2}$ of the binding energy, suggesting
hundreds of orbits for the transition from inspiral to merger.  The
period measured by an observer at infinity is $12 \pi \sqrt{6} M
\approx 90M$ for this orbit so that the decay time would be $\approx
10,000M$. For black holes of comparable mass, perturbation theory
cannot reliably treat the regime intermediate between an orbital
separation of $12M$ and merger.  However, as estimated in \cite{flan}
the decay time for this stage is $\approx 1500M$ and to join the
evolution smoothly to a post-Newtonian orbit at $20M$ would require an
evolution time of $\approx 10,000M$.

In our first test, we have successfully evolved an initially distorted,
moving (but non-spinning) Schwarzschild black hole for a time of
$60,000M$, clearly as long as needed for a smooth transition from the
post-Newtonian regime to merger, if this success could be duplicated in
the ultimate binary code.  The run was terminated because it had
achieved a steady state, with no sign of instability, and could be
extended further. It was carried out in a wobbling coordinate system
(see Fig.~3) which induces an ``artificial'' time dependence. (The
wobble of the outer worldtube $\Gamma$ in the vicinity of $r=7M$ is the
only time dependence seen at late times in the evolution). This
capability is important because it may not be possible to simulate a
binary in coordinates which become exactly stationary after the merger
and ring-down to final equilibrium.

Our second test establishes that the ingoing characteristic module
handles {\it spinning} black holes.  The outer world tube data is
induced from the exterior geometry of a Kerr spacetime with mass $M$
and angular momentum parameter $a=M/5$ (spin equal to $M^2/5$). The
metric is written in the Cartesian Kerr-Schild form~\cite{ks1}
\begin{equation}
       ds^2 = -dt^2 +dx^2 +dy^2 +dz^2 + 2H k_{\mu}k_{\nu}dx^{\mu}dx^{\nu},
\end{equation}
where $k_{\mu}$ is tangent to an ingoing congruence of twisting light
rays and $H$ is a potential (which equals $M/r$ in the $a=0$
nonspinning case).  The outer worldtube $\Gamma$ is located at
$x^2+y^2+z^2=49M^2$.  The module requires this worldtube data in Bondi
coordinates, which are spherical coordinates based upon the light cones
emanating inward from $\Gamma$ (advanced time coordinates). The
transformation from Cartesian Kerr-Schild coordinates to spherical
Bondi coordinates is carried out numerically in the neighborhood of the
worldtube by an extraction module which forms part of the Alliance's
Cauchy-characteristic matching procedure~\cite{cce,vishu}. The initial
data for a Kerr black hole on the ingoing light cone ${\cal N}_0$ is
complicated to specify analytically (the geodesic equation leads to
elliptic integrals). Instead, we choose initial data which approximates
Kerr data but distorts the black hole. The amount of distortion can be
measured in terms of the initial surface area of the MTS as compared
with the Kerr value ${\cal A}_K =8\pi M (M+\sqrt{M^2-a^2})$. In Fig.~3,
we plot ${\cal A}$ {\it vs} time for both the initially distorted Kerr
and wobbling Schwarzschild cases. Tests show that at late times
${\cal A}$ converges to the exact (Kerr or Schwarzschild) value as the
discretization size $\Delta \rightarrow 0$.

The Kerr evolution was run for a time of $15,000M$, at which the only
changes were at machine round-off. As apparent from Fig.~3,
the final Kerr equilibrium is effectively reached at $20M$.
 
The success of the ingoing characteristic module suggests a
possible strategy for excising the singularities in the binary case
(see Fig.~4). Two disjoint characteristic evolutions based upon ingoing
light cones are matched across worldtubes $\Gamma_1$ and $\Gamma_2$ to
a Cauchy evolution of the shaded region between them. The ingoing light
cones are each truncated at a MTS surrounding the singularities.  The
outer boundary $\Gamma$ of the Cauchy region is matched to an exterior
characteristic evolution based upon outgoing light cones extending to
infinity, where the waveform is calculated. This global strategy has
been successfully implemented for spherically symmetric
self-gravitating scalar waves evolving in a single black hole
spacetime~\cite{excise}.

Just as several coordinate patches are necessary to describe a
spacetime with nontrivial topology, an effective attack on the binary
black hole problem could be to patch together regions of spacetime
handled by different algorithms. The Cauchy-characteristic modules are
in place and calibrated for accuracy. The Alliance's Cauchy
module~\cite{all1} has evolved a Schwarzschild black hole for a time of
$475M$. Its performance and stability are now being studied in a wide
variety of tests. (For Cauchy evolution of a black hole with another 3D
code, see~\cite{seidel}). The key missing ingredient is the long term
stability of matching, which is a major current project.

This work was supported by the Binary Black Hole
Grand Challenge Alliance, NSF PHY/ASC 9318152 (ARPA supplemented).
Computer time was provided by the Pittsburgh Supercomputing Center.

\begin{figure}
\centerline{\epsfysize=80pt\epsfbox{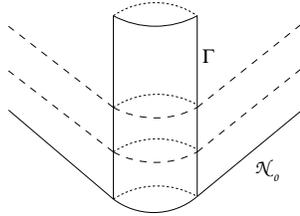}}
\caption{
The outgoing formulation: The exterior of $\Gamma$ is covered by a 
sequence of {\it outgoing} light cones.}
\label{fig:outg}
\end{figure}

\begin{figure}
\centerline{\epsfysize=80pt\epsfbox{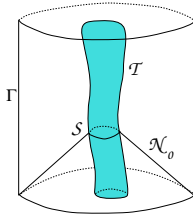}}
\caption{
The ingoing formulation: The interior of $\Gamma$ is covered by a sequence
of {\it ingoing} light cones. The interior of ${\cal T}$ is excised from the evolution.
}
\label{fig:ing}
\end{figure}

\begin{figure}
\centerline{\epsfysize=170pt\epsfxsize=180pt\epsfbox{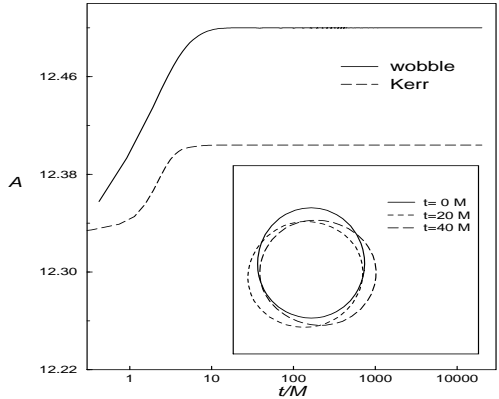}}
\caption{
Surface area {\it vs} time for a wobbling hole (with rotation
frequency $0.1$, offset $0.1$ and mass $0.5$) and an initially
distorted spinning hole (Kerr mass $0.5$). The inset shows three different
snapshots of the MTS in the case of the ``wobble''.
}
\label{fig:wobble}
\end{figure}

\begin{figure}
\centerline{\epsfysize=90pt\epsfbox{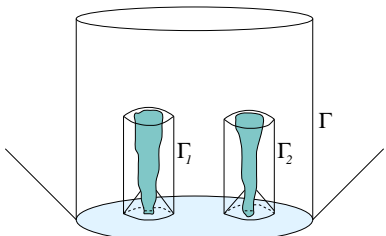}}
\caption{
A matching scheme for two orbiting black holes (in a co-rotating frame
which eliminates the major source of time dependence). 
}
\label{fig:blholes}
\end{figure}


\begin{references}

\bibitem{all}
For information about the goals and status of the Alliance visit:
{\bf http://www.npac.syr.edu/projects/bh/}.

\bibitem{all1} 
The Binary Black Hole Grand Challenge Alliance,
{\em Boosted 3-dimensional black hole evolutions with singularity
excision}, (submitted for publication).

\bibitem{all2} 
The Binary Black Hole Grand Challenge Alliance,
{\em Schwarzschild-perturbative gravitational wave extraction and outer
boundary conditions}, {\em Phys. Rev. Lett.} (to appear).

\bibitem{high}
N.T. Bishop, R. G\'omez, L. Lehner, M. Maharaj and J.  Winicour, {\em
Phys. Rev. D} {\bf 56}, 6298 (1997).

\bibitem{wobb}
R. G\'{o}mez, L. Lehner, R. L. Marsa and J. Winicour,
``Moving black holes in 3D'', submitted for publication,
gr-qc/9710138.

\bibitem{grail} 
S.L. Shapiro, S.A. Teukolsky, in {\em Dynamical Spacetimes and
Numerical Relativity}, ed. J. Centrella (Cambridge UP, Cambridge, 1986)
p. 74.

\bibitem{bondi}
H. Bondi, M.J.G. van der Burg and A.W.K. Metzner, {\em Proc. R. Soc. Lond. A}
{\bf 269}, 21, 1962.

\bibitem{penrose}
R. Penrose, {\em Phys. Rev. Lett.}, {\bf10}, 66 (1963).

\bibitem{gwsouth}
R. G\'{o}mez and J. Winicour,  in {\em Approaches to Numerical Relativity}, ed.
R. d'Inverno (Cambridge University Press, Cambridge, 1992) p. 143.

\bibitem{friedst2} 
H. Friedrich and J.M. Stewart {\em Proc. R. Soc. Lond. A}, {\bf
385}, 345 (1983).

\bibitem{papa}
R. G\'{o}mez, P. Papadopoulos and J. Winicour, {\em J. Math. Phys.}
{\bf 35}, 4184, 1994.

\bibitem{cce}
N.T. Bishop, R. G\'omez, L. Lehner, and J. Winicour, {\em Phys. Rev. D}
{\bf 54} 6153, 1996.

\bibitem{price}
R.H. Price, {\it Phys. Rev. D} {\bf 5} 2419, (1972).

\bibitem{jcp97} 
N.T. Bishop, R. G\'omez, P.R. Holvorcem, R.A. Matzner, P.
Papadopoulos, and J. Winicour, {\em J. Comput. Phys.} {\bf 136}, 236
(1997).

\bibitem{dinv}
M. Dubal, R. d'Inverno and C. Clarke, {\em Phys. Rev. D} {\bf 52}, 6868
(1995).

\bibitem{excise} 
R. G\'{o}mez, R. Marsa and J. Winicour,  {\em Phys. Rev. D} {\bf 56},
6310 (1997).

\bibitem{thornburg1987} 
J. Thornburg, {\em Class. Quantum Grav.} {\bf 4}, 1119 (1987).

\bibitem{pentrap}
R. Penrose, {\em Phys. Rev. Lett.}, {\bf14}, 57 (1965).

\bibitem{wald} 
R.M. Wald, {\em General Relativity} (University of Chicago Press,
Chicago, 1984).

\bibitem{hawkm}
S. W. Hawking, {\em J. Math. Phys.} {\bf 9}, 598 (1968).

\bibitem{flan}
\'{E}.\'{E}. Flanagan and S.W. Hughes, (submitted to {\em Phys. Rev. D}),
gr-qc/9701039.

\bibitem{ks1}
R.P. Kerr and A. Schild, {\em Proc. Symp. Appl. Math.} {\bf 17} 199 (1965).

\bibitem{vishu}
N. T. Bishop, R. G\'{o}mez, R. A. Isaacson, L. Lehner, Bela Szilagyi
and J.  Winicour, ``Cauchy Characteristic Matching'', in {\em On the
Black Hole Trail}, eds. B. Iyer and B. Bhawal (Kluwer, Dordrecht, to
appear).



\bibitem{seidel}
P. Anninos, K. Carmada, J. Mass\'{o}, E. Seidel, W.-M. Suen and
J. Towns, {\em Phys. Rev. D} {\bf 52}, 2059 (1995); P. Anninos, J. Mass\'{o},
E. Seidel and W.-M. Suen, {\em Physics World} {\bf 9}, 43 (1996).

\end{references}
\end{document}